\documentclass[letterpaper]{article} 
\usepackage{aaai25}  
\usepackage{times}  
\usepackage{helvet}  
\usepackage{courier}  
\usepackage[hyphens]{url}  
\usepackage{graphicx} 
\usepackage{natbib}  
\usepackage{caption} 
\usepackage{algorithm}
\usepackage{algorithmic}

\usepackage{newfloat}
\usepackage{listings}
\DeclareCaptionStyle{ruled}{labelfont=normalfont,labelsep=colon,strut=off} 
\floatstyle{ruled}
\newfloat{listing}{tb}{lst}{}
\floatname{listing}{Listing}
\title{\textit{Dead Zone of Accountability}: Why Social Claims in Machine Learning Research Should Be Articulated and Defended}
\author{
    Tianqi Kou\textsuperscript{\rm 1},
    Dana Calacci\equalcontrib \textsuperscript{\rm 1},
    Cindy Lin\equalcontrib \textsuperscript{\rm 2}
}
\affiliations{
    \textsuperscript{\rm 1}Pennsylvania State University\\

    \textsuperscript{\rm 2}Georgia Institute of Technology\\

}

\usepackage{bibentry}

\begin{document}

\maketitle

\begin{abstract}
Many Machine Learning research studies use language that describes potential social benefits or technical affordances of new methods and technologies. Such language, which we call ”social claims”, can help garner substantial resources and influence for those involved in ML research and technology production. However, there exists a gap between social claims and reality (\textit{the claim-reality gap}): ML methods often fail to deliver the claimed functionality or social impacts. This paper investigates the \textit{claim-reality gap} and makes a normative argument for developing accountability mechanisms for it. In making the argument, we make three contributions. First, we show why \textit{the symptom} - absence of social claim accountability - is problematic. Second, we coin \textit{dead zone of accountability} - a lens that scholars and practitioners can use to identify opportunities for new forms of accountability. We apply this lens to the claim-reality gap and provide a \textit{diagnosis} by identifying cognitive and structural resistances to accountability in the \textit{claim-reality gap}. Finally, we offer a \textit{prescription} - two potential collaborative research agendas that can help create the conditions for social claim accountability.
\end{abstract}

%

\section{Introduction}
Machine Learning (ML) research tends to make claims about potential social benefits or technical affordances of new methods and technologies. For example, McCradden et al. \cite{mccradden2023normative} have pointed out the lack of credible evidence in social claims about the efficacy or safety of ML-based medical care systems. In the domain of facial recognition, claims that ML models are “closely approaching human-level performance” \cite{taigman2014deepface} or that they represent “convenient and non-intrusive means for a wide range of identification applications” \cite{cheng2018surveillance} hype up the efficacy of these models. One stark counterexample is the false arrest of Robert Williams \cite{williams_2021} due to a facial recognition error, resulting in trauma for him and his family. Such cases demonstrate how ML models for facial recognition generally advertised as “convenient” or “reliable” have manifested the opposite of the claimed effects. Beyond individual cases, social claims about ML models’ capability can have even broader harms. For instance, ML studies claiming that \textit{Deep Neural Networks} can “detect sexual orientation from facial images” \cite{wang2018deep} or infer gender from handwriting \cite{bi2019multi} and facial images \cite{kumar2019gender} rest on reductive assumptions about sex, gender, and sexuality. By simply relying on these assumptions, these papers make implicit social claims. These flawed premises not only lack scientific rigor but also risk exacerbating systemic discrimination and psychological harm. These claims are socio-technical in nature but usually substantiated only by numerical evidence such as model performance.

Currently, there is a concerted effort within the AI ethics and ML community to develop accountability mechanisms for technology production. However, there is no accountability mechanism for when the advertised benefits or functionality in mainstream general-purpose or foundational ML research fail to manifest in practice. Namely, theorizing accountability for the communication of the impact of ML research that precedes technology production has thus far been largely ignored in the AI ethics community. In this paper, we shift the attention of AI accountability discourse from technology production to ML research production. We do this by identifying the reasons for this absence and providing potential solutions. In doing so, our analysis serves as a case study in applying the new accountability lens advanced in this paper.

\subsection{Outline}
\textbf{\textit{Section \ref{1} - Symptom}} describes the phenomenon of the \textit{claim-reality gap} and shows why the existing critical discourse needs new treatments to address it. Section \ref{1} also introduces the concept of the \textit{dead zone of accountability} as a more general phenomenon and a tool to explain why the claim-reality gap exists and why it continues to be resistant to accountability. \textbf{\textit{Section \ref{2}}} identifies three epistemic beliefs, and \textbf{\textit{Section \ref{3}}} the epistemological foundations of ML research, that both resist accountability in the \textit{claim-reality gap}. Sections \ref{2} on cognitive resistance and \ref{3} on structural resistance together \textbf{diagnose} that the \textit{claim-reality gap} constitutes a \textit{dead zone of accountability}. \textbf{\textit{Section \ref{4} - Prescription}} proposes two research avenues to create the conditions of possibility for accountability in the \textit{claim-reality gap}.

\section{Symptom: The Claim-Reality Gap and Dead Zone of Accountability}\label{1}
Deployed AI systems have led to abundant failures and harms \cite{kirchner2020automated, kirchner2020access, kippin2022covid,lecher2018happens}. The ML discourse is still hype-driven on both the building side and the critical side to different degrees. On the practitioners' side, overoptimism driven by computational innovation, building, and commercial profit has led to misuse and applications with little adequate testing. On the critical side, “Criti-hype” \cite{vinsel_2021}, a genre of AI critique that focuses on identifying and mitigating ML harms from models that "work too well" \cite{raji2022fallacy}, is detached from the reality that ML systems still often fail at basic tasks \cite{raji2022fallacy}. The grandiosity of language in hype-driven discourse allows the proliferation of exaggerated, unsubstantiated, and non-replicable claims with little to no evidential support \cite{kou2024model, birhane2022values, grill2022constructing}, creating a gap between claim and reality. As exampled in the introduction in \cite{mccradden2023normative, taigman2014deepface, cheng2018surveillance, williams_2021, wang2018deep, bi2019multi, kumar2019gender}, this gap is particularly stark when considering social claims - claims made in ML research that suggest social benefit or impact via ML \cite{kou2024model, birhane2022values}. In this section, we first scope the claim-reality gap and clarify what it applies to and how it appears. We then discuss why it is beneficial to distinguish \textit{the claim-reality gap} from the notion of \textit{socio-technical gap}. Finally, we define the dead zone of accountability and clarify its scope of application.

\subsection{Scope of the \textit{Claim-Reality Gap}} \label{1.1}
When we point to the \textit{claim-reality gap} as evidence of insufficient attention to social claims, we do not mean that such a gap exists in all ML research. For example, the Innovative Applications of Artificial Intelligence Conference (IAAI) has, for over three decades, fostered communities focused on “deployed applications with measurable benefits whose value depends on the use of AI technology.”\footnote{https://aaai.org/conference/iaai/} However, this orientation is not mainstream. While premier ML conferences such as NeurIPS, ICML, KDD, NAACL, and EMNLP include application-oriented tracks - e.g., NeurIPS’s “Applications,” “Machine Learning for Sciences,” and “Social and Economic Aspects of Machine Learning”; ICML’s “Application-Driven Machine Learning”; KDD’s “Applied Data Science Track”; and NAACL’s or EMNLP’s “Industry Track” - and while these venues make laudable efforts to develop high-quality benchmarks and datasets, papers in these tracks still tend to rely exclusively on benchmarking against standardized datasets. This is in stark contrast to IAAI papers, which typically provide more diverse supporting evidence for their social claims. Namely, despite a professed focus on application, the culture of treating model performance benchmarks as sufficient evidence still dominates ML research - yet such performance is far from enough to justify social claims (Kou 2024). Moreover, the \textit{claim-reality gap} applies even to studies that do not make explicit social claims, because, as we argue in Section \ref{2}, even “foundational research” focused solely on theory can bear implicit social claims that, in our view, should be made explicit.

\subsection{Why We Should not Subsume the \textit{Claim-Reality Gap} into the \textit{Socio-technical Gap}}\label{1.2}
One way to view the \textit{claim-reality gap}, and the social claims gap in particular, might be as a socio-technical gap, introduced by \cite{ackerman2000intellectual} as the space “between social requirements and technical feasibility”. Ackerman pointed out that human activities and social mechanisms are flexible, nuanced, and context-dependent but technical systems are usually rigid and unable to deliver flexibility and adaptability to reflect the social complexity of human activities. However, reducing the claim-reality gap to the socio-technical gap risks fostering two problematic tendencies. 

First, when critics use the socio-technical gap to make ML systems more accountable, it suggests that with enough societal inputs, ML practitioners' faulty social claims will improve. This ideology of "improvement" is pervasive in software work culture \cite{bialski2024middle}, and explaining the claim-reality gap as stemming from the socio-technical gap only emphasizes a "culture of improvement": the more societal or human factors are being accounted for, the more ML practitioners can close the socio-technical gap, and hence, the claim-reality gap. As such, subsuming the claim-reality gap into the socio-technical gap all but guarantees its continued invisibility. Subsequently, the claim-reality gap will self-perpetuate. Grandiose claims in the earlier days of deep learning research used to put out erroneous social claims using "suggestive language" have become pervasive in ML because of the self-referential nature of ML research culture and the tendency for junior scholars to cite the older work of senior scholars \cite{lipton2019troubling}.

Just like how earlier \textit{science and technology studies} writings of how laboratories extend out into the social world through acquiring allies, citations, and resources in support of their research \cite{latour1987laboratory}, the accumulation and dissemination of grandiose claims create a kind of historical legitimacy - ML scientists can cite previous papers’ social claims to gain epistemic authority and funding even if those benefits never materialized or had created the opposite effects. Even more worryingly, bolstering the legitimacy through the dissemination of false claims is turbo-charged given the speed and reach of ML research production.

Second, it implicitly endorses a division of labor where functionality and value alignment are addressed primarily at the design and implementation phase and considered a downstream responsibility by default (with the exception of deployment-focused ML research). Given these risks, we echo D’Ignazio and Klein’s question in considering data and feminism politics \cite{d2023data} - “Why should we settle for retroactive audits…if we can design with a goal of co-liberation from the start?” In the same vein, we ask: why should we settle for intervening at sites of design or deployment if we can also intervene at the sites of ML knowledge production - that is, where ML research is conducted and communicated before it is applied to the real world. This framing grounds our examination of the claim-reality gap and asks whether we can - and should - hold ML research itself accountable for the social claims it makes, even when it is further removed from applications than practitioners' work.

\subsection{Dead Zones of Accountability}\label{1.3}
Critical computing scholars have produced useful recommendations that could potentially improve engagement in social claims. \cite{green2020algorithmic} stated that computer scientists’ thinking during research design and evaluation is internal to the computational domain ("algorithmic formalism") and calls for “algorithm realism” and “porous thinking” to incorporate the broader context and fulfill their promises of enacting real-world impacts. Similarly, \cite{selbst2019fairness} identified five “abstraction traps” when designing fair socio-technical systems, another set of conceptual tools that computer scientists can utilize to address the claim-reality gap. Other examples include \cite{laufer2023optimization} where they reveal the normative commitments embedded in "optimization" in ML model training and \cite{hancox2021epistemic} where they discussed the contradiction between feature importance ranking (a common ML technique) and feminist values. Although socio-technical thinking has been adopted among some \cite{caton2024fairness, baker2022algorithmic, ehsan2021expanding}, they are still treated by most as external and rarely engaged in depth in mainstream ML research \cite{ashurst2022ai, ashurst2022disentangling, kou2024model}. For example, in subareas of ML such as transparency and interpretability, the framing becomes over-technical and tends not to serve the public \cite{corbett2023interrogating}.

As we will show in Sections \ref{2} and \ref{3}, the sparse adoption of critical computing perspectives (resulting in the \textit{claim-reality gap}) in mainstream ML research production is not coincidental but reflects what we term a \textit{dead zone of accountability} which we define in the context of ML research as:
\begin{quote}
    \textit{aspects of the ML ecosystem - such as methodologies, tools, institutional norms, evaluation metrics, and power structures - that resist critical scrutiny and meaningful accountability.}
\end{quote}

Although our conceptualization of \textit{dead zone of accountability (DA)} is derived from \textit{dead zone of imagination (DI)}, there are differences we want to highlight. To do that, we synthesize the definition of DI \footnote{Graeber never gave an explicit definition} to be the following:
\begin{quote}
    \textit{aspects of bureaucratic or institutional practices that appear mundane or meaningless on the surface but conceal deep power imbalances and structural violence.}
\end{quote}

Unlike \textit{DI}, which explicitly highlights structural violence and power imbalance, \textit{DA} does not explicitly require structural violence to be present. This distinction is crucial when identifying a \textit{DA}. Accountability is a normative concept presupposing rules, standards, and responsibilities; resistance to accountability (which characterizes a \textit{DA}) necessarily implies a power imbalance, since the legitimacy of the status quo and distribution of harms or benefits are contested. Therefore, thorough analyses of a phenomenon as a \textit{DA} should incorporate, or be followed by, an examination of power dynamics. This paper does both for the \textit{claim-reality gap}: characterizing it as a \textit{DA} (Sections \ref{2} and \ref{3}) and revealing its underlying power imbalance (Section \ref{1.3}).

\subsection{\textit{Structural Violence} in the \textit{Claim-Reality Gap}}\label{1.4}
Graeber's analyses of the practice of filling in paperwork showed that healthcare seekers must navigate complex rules and manage cross-institutional communications to avoid being denied necessary care. Meanwhile, bureaucrats are not required to reciprocate this level of effort or understanding/imagination. "Filling paperwork" - may appear devoid of deeper meaning but, upon closer analysis, reveal underlying power imbalances. Specifically, the distribution of physical, emotional, and imaginative labor disproportionately burdens those disempowered, often with the threat of \textit{structural violence} - losing access to health care.

The actors involved in the \textit{claim-reality gap} typically include two sides. First, the ML research community - the "coding elites" \cite{burrell2024automated} - with access to vast research funding and significant economic and social influence. Second, the impacted communities, regulatory agencies, think tanks, civic tech activists (such as technology auditors and investigative journalists), AI ethics scholars, and Human-Computer Interaction researchers. These communities work tirelessly to identify harms, uncover their causes, and propose solutions, including legal regulations, conceptual frameworks, tools, and community organizing to mitigate the damage caused by technologies rooted in the ML community’s knowledge products. \cite{phan2022economies} For many, this effort constitutes their primary career or professional focus.

Communities outside of ML research must acquire technical literacy and navigate an unfamiliar world to understand and address the threats partially constituted by ML knowledge. Meanwhile, ML scientists-except for a rare minority who engage in cross-disciplinary or downstream work-rarely reflect on the experiences and efforts of those striving to make technologies function and/or less harmful. Their focus remains on unrealistic benchmarks detached from real-world contexts \cite{raji2021ai}.

As David Graeber observed: “[T]he process of imaginative identification [is] a form of knowledge … within relations of domination, it is generally the subordinates who are … relegated the work of understanding how the social relations in question really work” \cite{graeber2012dead}. This illustrates a form of structural violence reflected in the current division of interpretive labor in ML research. Communities outside of ML research must exert tremendous efforts to understand the technical construction of and undertake socio-technical analysis of ML methods and technologies and develop solutions, otherwise they are subjected to harms that are inextricably connected to general-purpose ML research.

Although this uneven distribution of interpretive labor is not unique to ML and can be seen in other scientific disciplines, it is particularly pronounced in ML research. Fields such as medicine often involve longer vetting and maturation periods before their knowledge has real-world impacts. In contrast, ML research frequently bypasses iterative testing and expansive evaluation, rapidly transitioning from being "exploratory" \cite{bouthillier2019unreproducible, herrmann2024position} claims to real-world applications due to the low barriers to entry and the accessible computational resources.

Graeber made a moral case for revealing and addressing structural violence. To illustrate this point, Graeber uses a metaphor: a master whips a slave to enforce “unquestionable obedience” and establish “absolute and arbitrary power.” Once obedience is achieved, the master can remain completely oblivious to the slave’s experience, while the slave must deeply understand and navigate the master’s perception to survive. Graeber argues that scholars have a moral responsibility to interrogate “what ‘unquestioning’ actually means” and by extension, unveil the power dynamics between master and slave. Similarly, the structural violence this paper identifies propels scholars to identify zones such as the \textit{claim-reality gap} and pose questions about the legitimacy of its existence.

Put differently, in the context of ML research and those subjected to their social claims, scholars critiquing such systems should also question (and not simply obey) not only the outcomes of these systems (as many have in the fields of algorithmic fairness and accountability) but also the social claims researchers make of the world. 

\subsection{Application of \textit{Dead Zone of Accountability}}\label{1.5}
The broad phenomenon of the \textit{claim-reality gap} is one example of a dead zone of accountability. We envision dead zones of accountability as applicable at both local levels (such as individual studies or subfields) and broader disciplinary contexts. Using this paper as an example, we identify the \textit{claim-reality gap} as a \textit{dead zone of accountability} by demonstrating that it is sustained by cognitive or cultural barriers - beliefs and assumptions (Section \ref{2}) - and structural or procedural barriers (Section \ref{3}) that are “hostile” to accountability within this gap.

In this paper, we are primarily concerned with revealing the \textit{conditions of possibility} for accountability in the \textit{claim-reality gap}\footnote{We use conditions of possibility in the Foucauldian sense \cite{foucault2005order}. In his archaeological analysis of knowledge, Foucault uses \textit{condition of possibility} to refer to the implicit rules, discursive formations, and epistemic constraints that shape knowledge production.}, and how the conditions of possibility thus far foreclose new and alternative forms of accountability in ML research.\footnote{Existing accountability mechanisms primarily target actors directly involved in the building and deployment of ML systems. By new and alternative, we refer to mechanisms such as social claim accountability that could link upstream ML researchers more closely to the real-world contexts in which their work may be used. At present, "accountability measures" for those producing general-purpose ML research typically take the form of impact statements or ethics checklists - tools that have been shown to be inadequate and lacking in critical depth \cite{ashurst2022ai, kou2024model}; more importantly, these measures are loosely enforced, recommended but not required} We leave tracing causal links between social claims and harm, and detailed recommendations of how to improve engagement with social claims, to future work. 

By focusing on the \textit{conditions of possibility}, we suggest that there are untapped forms of accountability that could emerge if ML research engaged with social claims more seriously. To create these conditions of possibility, we argue that AI ethics and ML research must challenge these limiting beliefs (in Section \ref{2}), resist \textit{computational capture} (in Section \ref{3.1}), and strive toward a new form of epistemic authority in (Section \ref{3.2}).

Through the lens of \textit{dead zone of accountability}, critical computing scholars and practitioners can pose questions about processes, actions, and entities that traditionally appear unaccountable - whether practically or normatively - such as the links between concrete harms and research culture, methodological practices, or computational tools and techniques. These inquiries offer valuable contributions to \textit{the responsibility gap} discourse, which focuses primarily on tracing harms to stakeholders \cite{goetze2022mind, kou2024model}. Labeling something a dead zone of accountability does not simply mean the absence of accountability mechanisms; it requires revealing why accountability is resisted and what practical or epistemological obstacles prevent it.

\section{Cognitive Resistance to Accountability}\label{2}
In this section, we identify three flawed epistemic assumptions that make the \textit{claim-reality gap} resistant to accountability.

\subsection{“Making ML scientists Account for Social claims Misunderstands Their Role Responsibilities”}
One presumption may be that it is not within the “role responsibility” \cite{douglas2009science} of ML scientists to produce well-justified social claims. As Douglas explains, “Role responsibilities arise when we take on a particular role in society, and thus have additional obligations over and above the general responsibilities we all share.” \cite{douglas2009science} In the public and academics' imagination, the primary responsibility of ML scientists’ role is to develop generalizable models that parallel or surpass human learning and problem-solving capabilities. Success in this role is measured through computational performance. Additional responsibilities, such as ensuring transparency, replicability, and objectivity, are framed as epistemic virtues, which are typically also operationalized computationally, through practices like sharing code and data, enabling others to replicate studies and reproduce model performance.

Under such a presumption, two corollary justifications might follow. First, asking ML scientists to take accountability for speculative social claims is seen as exceeding their role responsibilities because ML scientists, similar to other scientists, “are not trained to deal with social and political issues” \cite{pamuk2024politics}. As such, ML social claims are often dismissed by "experienced" readers - those familiar with widespread replication and functionality failures and value-alignment issues - as frivolous, unreliable, or erroneous. Social claims only attract scrutiny when they are tied to concrete, harmful applications, prompting actions from “fact-checking” communities such as auditors, regulators, civic tech activists, and critical computing scholars. Examples include the analysis of facial recognition technologies and algorithmic decision-making systems aimed at mitigating harm and improving functionality.

However, it is a mistake to assume that “experienced readers” represent the entirety of the audience of speculative social claims with widespread misuse and unwarranted hype.  Similarly, accepting the status quo scope of "what they are best equipped to do" as sufficient and acceptable is equally flawed. On this note, it is crucial to recognize that when it comes to justifying social claims there is a middle ground between \textbf{\textit{doing it all}} and \textbf{\textit{doing nothing at all}}. Namely, this paper's stance on accountability about social claims is not asking ML scientists to become social scientists or domain experts.

The second corollary justification under this presumption is: that the nature of scientific knowledge - being incomplete, uncertain, and fallible - means that ML scientists often cannot predict who will use their models or for what purpose. As such, regulating social claims through accountability is seen as infringing the nature of knowledge and, therefore can cause the epistemic harm of limiting the freedom of inquiry - a normative principle tied to the postwar funding ethos of science, which holds that unfettered scientific exploration benefits society collectively \cite{bush2021science, douglas2024social}. Under this view, scoping ML scientists' responsibility within the computational domain and unburdening them with social reflections is deemed efficient and fair, allowing them to focus on what they are best equipped to do while preserving epistemic freedom \cite{douglas2024social}.

However, it would be misguided to assume that regulating social claims will come at the expense of limiting free inquiry. History provides numerous examples where scientific progress has benefited from external social pressures on scientists' conception of the social world. One salient example is \cite{epstein1996impure}'s ethnographic and sociological study on AIDS activists' impact on knowledge production in the scientific and medical communities. Amidst the AIDS crisis in the 1980s, AIDS activists contested the "widely accepted scientific approach...to design trials" \cite{pamuk2024politics} - using homogenous populations disregarding sex and race. The activists rejected the use of homogenous populations as ill-suited to fulfill the practical purposes of AIDS research. As a result of social reflection, scientists developed two scientifically valid methodologies - the "pragmatic approach" that was aligned with clinical practices and the conventional "fastidious approach" \cite{epstein1996impure}. This example highlights how social considerations can be conducive to methodological diversity. To harness these benefits, we need to have suitable institutional structures or norms within science that effectively define and "administer" social considerations.

\subsection{“Easy Solution! Let's Erase Social Claims in ML Papers”}
Some readers might view our stance on regulating speculative claims as excessive and propose a simpler solution: remove all social claims and any promissory tone from both applied and foundational ML papers and leave nothing to regulate. However, even if explicit social claims or subtle allusions to social benefits - whether regarding functionality or upholding social values - are eradicated from written texts, social claims will still be implicitly made. To evaluate the feasibility of erasing social claims, one must contextualize this proposal within ML culture. ML research does not exist in a vacuum; it operates within a complex network of communities, institutions, and relationships \cite{widder2023dislocated, bennani2024infrastructuring}. Currently, the ethos of ML culture is grounded in general applicability, functional utility, and "techno-optimism" \cite{campolo2020enchanted}. These foundational aspects of ML culture cannot be extinguished by merely removing textual expressions of social claims. 

Over the past decade, research agendas centered on tech ethics and social good have gained prominence in AI ethics and premier ML conferences. \cite{phan2022economies} For example, the 3.77\% of NeurIPS submissions flagged for an ethics review in 2023 are bucketed into categories specifically concerned with concrete harms such as bias, privacy violations, and research integrity issues. \cite{mackey_poland_chen_2023} This academic environment has fostered implicit expectations that researchers must navigate to secure funding, recognition, and citations. Researchers are now evaluated not only on their capacity for technical innovation (through formal peer review processes) but also on the moral valence and practical utility of their work (through general culture). Put simply, it is sexy and even extremely rewarding for ML researchers to create AI not only for social good but also for the social world. 

This dual expectation of technical contribution and social impact has introduced two criteria through which ML research is interpreted and evaluated - first, is said ML research technically innovative? Second, is said ML research socially impactful? A published study might omit explicit references to social benefits, yet readers are likely to assume the work has the potential to deliver utility and/or ethical impact. For instance, tenure-track faculty in ML and AI tracks are evaluated on how impactful their research is. \cite{patterson_snyder_ullman_1999} As a result, social claims - whether explicit or implicit - become deeply embedded in a hype-driven culture where both ethical alignment and practical utility are deeply intertwined and rewarded. Thus, the idea of fully disentangling and removing social claims from evaluations of a successful ML research output is not feasible because social claims can be inferred by reviewers, funders, or the public.  Take ML-based science as an example (ML models or methods are introduced in other sciences to develop understandings of a phenomenon), the lack of awareness about the epistemological limitations of ML in scientific inferences \cite{humphreys2020automated} has led to the production of "pseudoscience" \cite{andrews2024reanimation} and "inaccurate findings". People tend to take claims and numbers as "granted" \cite{grill2022constructing, raji2022fallacy}. 

Thus, the suggestion to erase social claims from ML papers is neither feasible nor productive. Such an approach ignores the shifting cultural realities of the field, and how CS and ML professions are rewarded for making social claims. It also allows social claims to remain unacknowledged and unregulated while failing to address the negative impacts of its circulation and uptake.

\subsection{“This is Too Hard, Let Practitioners Worry”}
The third justification is grounded upon a sense of pessimism about developing accountability mechanisms for social claims. Even those sympathetic to the idea of regulating speculative claims often struggle to envision how this could be achieved.

Despite the added socio-political and ethical dimensions, accountability fundamentally rests on qualities and actions central to accounting - recording, documentation, and traceability \cite{cooper2022making}. To explore this further, consider \cite{bovens2007analysing}'s (2007) conception of accountability - “a relationship between an actor and a forum, in which the actor has an obligation to explain and to justify his or her conduct, the forum can \textit{pose questions} and \textit{pass judgment}, and the actor may \textit{face consequences}” (italic added). Boven’s \cite{bovens2007analysing} conception of accountability highlights answerability to harm as an essential component of accountability relationships.

The pessimistic view arises from the undeniable reality that we currently lack a clear framework for how ML scientists, trained to produce generalized models, should engage with the questions and demands of other social groups.  There is no established direct and answerable “relationship” between the actor (ML research community collectively) and the forum (those impacted, positively or negatively, by their knowledge). This is unlike the extensively investigated relationship between tech designers and impacted communities under the Participatory AI movement \cite{young2024participation, delgado2023participatory}. Consequently, the essential elements of accountability - \textit{posing questions}, \textit{passing judgments}, \textit{addressing harm}, and \textit{facing consequences} - are absent. ML scientists form a closed expert community that communicates in the computational domain, distinct from the language used by communities who articulate lived experiences.

Answerability implies an affinity toward being able to trace the causal chain from an entity/action to harms and having conceivable remediative actions (usually expressed as stakeholder actions and responsibilities) at the junctures of the causal chains - auditing of, legal regulation on, and participatory design of, technologies, etc. However, when moving upstream from ML technology to ML knowledge, we enter un "uncharted" territory - the pathway from speculative and generic social claims to tangible failures and harms becomes even more nebulous.

This lack of answerability and traceability often leads the above-characterized pessimism to morph into a normative argument against regulating speculative claims. Some may assert that "knowledge is just knowledge" and that “dangers and ethical issues only arise when science is applied in technology” \cite{wolpert1999science}. However, decades of analysis from science studies scholars like \cite{harding2013rethinking} and \cite{keller1982feminism}, etc., constituting counter-evidence to this stance, highlighting the inherently social nature of scientific knowledge and practice.

While we may not yet understand the precise pathway from a single social claim to concrete harm (unlike tracing harm to a specific technological design), this uncertainty does not justify inaction. Pessimism rooted in the unknown is merely intellectual laziness. Just because something is currently unknown does not mean it is unknowable - or that we should not attempt to understand it.

These three (likely non-exhaustive) faulty yet pervasive assumptions collectively position ML scientists within an "epistemic bubble" \cite{douglas2024social}. By framing their primary contributions as computational and generalizable - detached from specific contexts or applications - ML scientists are afforded considerable freedom in crafting speculative claims, including their content, validity or whether or not to make them explicit. This freedom effectively absolves them of responsibility for the implications of these claims. With such an epistemic bubble encasing the dead zone of accountability, ML scientists continue to gain epistemic authority and funding through speculative social claims. Meanwhile, downstream communities, frustrated by the chaos and harm perpetuated by unwarranted ML hype, find themselves unable to contest the legitimacy of these claims. Instead, they are forced to accept the status quo as an unfortunate byproduct of the dead zone. These three popular epistemic assumptions are partially why ML researchers have not attempted to rectify the issue of absence of accountability in the \textit{claim-reality gap}. The next section discusses the disciplinary logics of ML research that reinforces these beliefs. 

\section{Structural Resistance to Accountability}\label{3}
In this section, we introduce one key epistemological underpinning of ML research that causes its systematic resistance to accountability:  \textit{computational capture}. We define \textit{computational capture} in the field of ML as:
\begin{quote}
    \textit{processes through which computational goals exclude or sideline social or practical goals through research practices, domain norms, or incentives structures}
\end{quote}

In Section \ref{3.1}, we illustrate at least two ways in which \textit{computational capture} manifests: through methodological rigidity and techno-elite control. In Section \ref{3.2}, we explore how ML's current form of epistemic authority relates to these manifestations in ways that further reinforce computational capture.

\subsection{Two Instances of \textit{Computational Capture}}\label{3.1}
\subsubsection{4.1.1 Methodological Rigidity as \textit{Computational Capture}}
Computational capture can manifest in how problems are selected, studies are designed, tools are chosen, and success is measured, ultimately regimenting research into patterns that prioritize “SOTA-chasing” \cite{church2022emerging}. As a result, the ML research process is characterized by the Common Task Framework (CTF) \cite{liberman2010obituary}. \cite{donoho201750} defined the CTF to have the following ingredients: (a) a publicly available training dataset, (b) a set of common tasks, and (c) a shared evaluation mechanism.

Under \textit{computational capture}, the field tends to deem only "general" knowledge acceptable for the discipline, prioritizing justifying decontextualized model performance over studies connected to localized applications \cite{karl2024position, wagstaff2012machine, henderson2018deep}. This limits or even punishes, methodological pluralism. For example, in the autoethnographic study by \cite{rubambiza2024seam}, where computer scientists (computer network research more specifically) attempting to deviate from computational framing of innovation faced repeated rejections of their submissions as irrelevant and trivial for the field. This caused them to compromise their vision and presentation of “societal impact”. Similarly as \cite{leonelli2023philosophy} pointed out: “Researchers working [within] existing \textit{[research] repertoires} \cite{ankeny2016repertoires} may not value - or even consider - elements that are not already part of that repertoire…when wanting to modify a \textit{[research] repertoire}, researchers may face significant hurdles - in the shape of negative reviews, rejection by funding bodies and critical questioning by powerful peers.”. 

The incremental and minuscule improvements in model performance made over shared tasks are proven to be unstable \cite{karl2024position}. Even when they are stable, these models fail to emulate human intelligence, often struggling with the most basic real-world tasks. General purpose ML research's state mirrors \cite{mccarthy1997ai}'s criticism of using computer chess as evidence of programs emulating human intelligence: “It was as if geneticists had focused their research efforts on breeding Drosophila (fruit flies) to race against each other… [w]e would have some science, but mainly we would have very fast fruit flies.” One might invoke Thomas Kuhn’s \cite{kuhn1997structure} conception of the scientific revolution and argue that ML research is simply in the phase of "normal science", and a paradigm shift will occur with time. However, as Paul Feyerabend \cite{feyerabend1970consolations} pointed out, a revolution requires knowledge and inquiries that challenge the presiding paradigm.

In this regard, methodological capture not only limits meaningful progress in innovation, aggravating the creativity statement \cite{cho_2024, hill_2024}, but also restricts the room for methodological pluralism that integrating social considerations requires.

\subsubsection{4.1.2 Techno-elite Control as \textit{Computational Capture}}
Techno-elites refer to powerful actors in ML ecosystem industry and academic ML research labs, funding bodies across industry and federal agencies. These actors' visions drive actions (e.g. assigning funding, setting conference aims and organization, determining publication and review standards, defining methodological rigor) that draw epistemic boundaries - centering ML research focused on producing state-of-the-art model performance on shared benchmarking datasets over meaningful engagement with social, ethical, or interdisciplinary considerations.

Not only does the computational capture constrain ML within computational formalism, but it also betrays how the funders of ML research, who are "close to the source of knowledge" (mostly privatized funding from tech giants\cite{whittaker2021steep}), monopolize the research paradigm for the preservation of existing paradigm and their benefits. This reality casts doubt on the prospects for a disciplinary revolution in ML research.

With the highly standardized and mathematized ML research process's main beneficiaries being the ML community, it aggravates the misalignment between the producers of ML knowledge and those impacted by it. Different communities have disparate assessments of the acceptability of uncertainty of knowledge \cite{pamuk2024politics}. ML scientists think of the variability of a model's performance across context or training iterations as an inherent statistical property whereas other communities prefer empirical evidence to give real-world meaning to such uncertainty. The choice of handling uncertainty requires not just an epistemic judgment, but also a moral one. Under computational capture, the decision over what is the right interpretation of this uncertainty is made by techno-elites. For example, misclassification of a few or a few thousand instances might have no impact on the model performance, leaving ML scientists unbothered, which can materialize into devastating and traumatizing effects in real life, as demonstrated in the case of Robert Williams \cite{williams_2021}.

\subsection{Inadequate Epistemic Authority Under \textit{Computational Capture}}\label{3.2}
Under computational capture, ML scientists’ epistemic authority rests on their expertise in designing models and evaluating performances against benchmark datasets. This subsection shows that such authority is increasingly insufficient for reflecting the diversity of current ML research. In today’s context, a complementary form of epistemic authority - centered on practical efficacy - has become urgent. ML scientists should share the task of cultivating this new form of epistemic authority with practitioners who work at the sites of system building and deployment.

Science and Technology Studies scholars \cite{haraway2013situated, harding2013rethinking} showed that objects of knowledge in science are historically constituted. The same object of study is represented differently across time, attributed varying scientific or cultural meanings, and associated with different effects. For example, objects of study (such as “rocks” and “electrons”) in the natural sciences, though intuitively perceived to possess permanence and stability, are “socially constituted” \cite{harding2013rethinking}. Without delving into whether ML is science or engineering, it is crucial to stress that objects of inquiry in ML - usually mathematically constituted - are also socially shaped. This has broad implications for research culture beyond the objects of inquiry themselves.

ML has its early roots in AI research from the 1950s. At the time, computer scientists’ artifacts were basic programs and rule-based games like chess, and the field held a scientific register with focuses on intelligence, learning, and understanding - qualities associated with Artificial General Intelligence (AGI). AI scientists in the earlier decades hoped that computer programs could demonstrate these qualities \cite{minsky1961steps, o2013giants, bennani2024infrastructuring}. Current day ML research submissions also serves "scientific, engineering, and societal goals" \cite{blili2025stop} which can be observed from ML conferences' submissions, and usually these goals are mixed \cite{blili2025stop}. It is important to note that even the most "basic" or "foundational" ML research is \textbf{not} \textit{application-neutral - produced with no knowledge or intention for how it might be used}. ML scientists must respond to funders’ needs to accrue funding as a community, therefore, unavoidably \textit{not} application-neutral. For example, an analysis of U.S. Department of Defense (DoD) grant solicitations targeting AI researchers \cite{widder2024basic} reveals how ML scientists and the DoD engage in mutual enlistment, and how “basic research” often obscures the fact that these projects serve militaristic agendas. The privatization of ML research by tech giants also indicates that basic research responds to research agendas set by private funders \cite{whittaker2021steep}. Put simply, AI has evolved from being “chiefly a scientific research subject during the twentieth century to a widely commercialized, [\textit{politicized, and weaponized}] innovation” [italic texts are added] \cite{liu2021sociological} and it serves a broad set of goals that in ways render ML research \textit{tools} rather than only scientific understanding.

The prominent goals of ML for science, technology, and social benefits call for a form of epistemic authority centered on practical efficacy, different from epistemic authority over model building and evaluating in lab settings. In other words, “predictive performance alone is not a good indicator of the worth of a [ML] publication” \cite{karl2024position}. However, one question remains, why should ML scientists cultivate this new form of epistemic authority instead of just practitioners who "specializes" in actualizing concrete goals. 

Epistemic authority over practical efficacy means the capacity to adequately evaluate, explain, and justify how well an ML study serves the purported goals with considerations of intended contexts, operational constraints, and emergent society impacts, etc. There are two reasons for why ML scientists should cultivate this form of epistemic authority. First justification is ethical. The knowledge mainstream ML research produces is general but comes with a \textit{generality debt} - "relying on the generality or flexibility of tools to postpone crucial engineering, scientific, or societal decisions" \cite{blili2025stop}; this brings us back to Section \ref{1.4}, the postponing of difficult decisions will exacerbate the structural violence that is already great. Second justification is practical. ML scientists have unique influence over the choice of projects, model architectures and evaluation methods, infrastructural requirements, and the discursive dimension - shaping narratives and pre-empting misuse.

Sections \ref{2} and \ref{3} together are our diagnosis for the absence of accountability in the \textit{claim-reality gap}, and the diagnosis indicates that both epistemic assumptions and ML research regime resist accountability, and these resistances render the \textit{claim-reality gap} a \textit{dead zone of accountability}. The final section will offer potential solutions to create conditions for developing accountability in the \textit{claim-reality gap}.

\section{Prescription: Social Claims Should be Articulated and Defended}\label{4}
This section outlines strategies to rectify the conditions that we identified throughout the paper to create possibilities for accounatibility in the \textit{claim-reality gap}. In Section \ref{4.1}, we identify a barrier that can discourage epistemic reform in ML research and connect this to the discussion in Section \ref{3.2} on epistemic authority as a means to mitigate such discouragement. In \ref{4.2}, we propose solutions to cultivate ML scientists with a new form of epistemic authority and create the conditions necessary for accountability.

\subsection{Cultivating a New Form of Epistemic Authority}\label{4.1}
ML scientists risk certain losses if their research is held accountable for social considerations. Epistemically, their authority would be subjected to unfamiliar standards and contestation beyond relatively narrower computational evidence. Materially, justification of societal impact would face greater scrutiny, possibly threatening funding opportunities and diminished socio-economical status. As Porter \cite{porter1996trust} observed:

\begin{quote}
    “[S]cience is [made by] communities that are often troubled, insecure, and poorly insulated from outside criticism. Some of the most distinctive and typical features of scientific discourse reflect this weakness of community…[O]bjectivity in science is…partly a response to the…pressures.” [p. 230]
\end{quote}

The rigidity and control - characteristic of computational capture - serve to generate a shared discourse that unifies a "weak research community". These boundaries demarcate the field to resist external contestation, but engagement with the social breaks down these boundaries and incorporating knowledge from other domains can indeed be destabilizing \cite{burrell2024automated}.

Green and Viljoen (2020) argue that reform demands computer scientists recognize their positionality and develop the epistemic capacity to navigate sociopolitical dimensions. While we share this view, we emphasize that many ML scientists - especially those working on general-purpose models - care about social good, but do not primarily identify with the ML-for-social-good community as their main focus. Translating critical insight into systemic reform means acknowledging this heterogeneity. Reform efforts must recognize and accommodate this "epistemic diversity" within ML community\footnote{In the context of the sciences, Leonelli \cite{leonelli2023philosophy} identifies six broad categories to understand scientists' epistemic diversity: material, conceptual, methodological, infrastructural, socio-cultural, and institutional.}.

Given the disciplinary "insecurity", any effort to institutionalize socio-technical responsibility must acknowledge the pushbacks these reforms might face due to potential loss of resources and diminished epistemic authority. For this reason, we propose the \textit{rigorous articulation of social claims} as a strategy likely to encounter less resistance as it affords ML scientists a new form of epistemic authority. 

If social claims become the basis for a new form of epistemic authority, challenging computational capture would expand rather than diminish ML scientists’ authority. Current reform tools - such as impact statements and ethics checklists - often cast sociotechnical concerns as limitations to be disclosed \cite{ashurst2022ai}. In contrast, when ML scientists explicitly ground their study in potential contexts, reason with limitations, uncertainty, and bias, they demonstrate the kind of intellectual rigor that defines good science. Efforts like these should be recognized and rewarded. Establishing this new epistemic authority means that computation alone does not determine the relevance or value of ML studies or the competence of ML scientists. Instead, they are evaluated for the quality of their reasoning and their ability to articulate both computational and social dimensions as claims - not simply for producing high, but often ambiguous, model performance scores \cite{blili2025stop}.

The new form of epistemic authority mentioned requires specific arrangements in ML research practices. Epistemic authority is derived from expertise; assigning epistemic authority over practical efficacy implies expertise in proving practical efficacy. Just as model performance claims go through rigorous scrutiny, we propose that social claims should also undergo scrutiny. This brings us to the suggestion that social claims should be treated as knowledge claims that need to be articulated and defended with evidence, rather than being speculative or implicit. These knowledge claims should be justified with "sound reasoning and credible evidence". If a shared standard for such reasoning emerged, we could potentially design accountability mechanisms to hold ML researchers responsible for not specifying their social claims and/or for not providing sufficient evidence for them. Reform of this kind is within the would be internal to ML research instead of compartmentalizing tasks of computation and social in a separate basket such as ethical checklists or impact statements.

\subsection{Two Collaborative Research Agenda of Developing Conditions for Social Claim Accountability}\label{4.2}
The goal suggested at the end of 4.1 is not easy to achieve because the maturing and reform of scientific norms and practices takes time. The Human-Computer Interaction (HCI) community has a history of developing tools or mechanisms to improve the rigor of scientific research, such as \cite{feger2019role}. In this section, we propose two collaborative research avenues that involve the HCI community to develop and implement evaluation mechanisms for social claims to create conditions of possibility for social claim accountability mechanisms.

\subsubsection{5.2.1 Identify Categories of Social Claims}
ML papers frequently make broad social claims without specifying concrete applications. This vagueness is partly intentional, as it implies the generalizability of the research findings while acknowledging the difficulty of predicting all possible future uses of a method. The uncertainty around the balance between generality and specificity poses a challenge to mainstream ML researchers who focus on benchmarking performances - they may struggle to scope social claims appropriately or to assemble suitable supporting evidence. While ML researchers are generally clear on the standards for justifying a computational contribution, there are few established norms or examples for substantiating more complex, context-dependent social claims.

To bridge this gap, a critical first step we suggest is developing an understanding of social claims by aggregating existing social claims from ML papers and deriving social claims from ML domain applications with the help of domain experts and other civic tech communities. Eventually, the social claim research community can 1) help develop social claim categories with case examples and 2) derive commonly used evidence to support different categories of social claims. This growing repository of social claims with case examples, and commonly used evidence (for each type of claim) can serve as pedagogical resources that can be integrated into ML scientists' training for socio-technical understanding of their work, and as a reference tool to substantiate their social claims when evaluating their methods and drafting manuscripts.

\subsubsection{5.2.2 Help ML Researchers Systemize Supporting Evidence}
ML research workflow is characterized by the Common Task Framework \cite{donoho201750} which emphasizes standardization of research process and reporting. However, such standardization is not observed in evidence presentation. Even for model performance claims - arguably the most rigorously evaluated aspect of ML research - there is typically no explicit representation of the relationship between a claim and its supporting evidence. Reviewers mainly rely on their trained judgment to evaluate its validity. However, this lack of transparency can obscure the reasoning process. To address this gap, HCI researchers can collaborate with ML researchers to identify common evidence in ML research. Common forms of evidence include benchmarking comparisons, cross-validation, ablation studies, A/B testing, deployment observations, and field feedback \cite{kou2024model}.

Besides identifying key types of evidence, the HCI and ML communities could co-develop tools for documenting, organizing, and presenting evidence in ML papers. For instance, one approach could involve the use of diagrams that map individual claims to their supporting evidence, with each diagram representing a single claim. These visualizations could illustrate the logical connections between different types of evidence and the claims they support, offering a clear view of the study’s inferential structure. Beyond improving clarity for fellow researchers, these tools would make ML research more accessible to less technical audiences, such as policymakers, AI auditors, and domain experts, who lack the specialized ML knowledge needed to parse technical content.

\section{Conclusion - "Love Your Monsters"}
This paper coins the term \textit{dead zone of accountability} - a lens through which scholars can identify aspects of ML ecosystem that resist accountability. We demonstrated that our target case - the \textit{claim-reality gap} - is a dead zone of accountability by showing the epistemic assumptions and ML epistemological underpinnings that resist accountability for \textit{claim reality gap}. We ended the paper with two collaborative research avenues that could create the conditions for developing accountability mechanisms for the \textit{claim-reality gap}.

The AI ethics discourse underscores the importance of stepping out of \textit{technological somnambulism }\cite{winner1983technologies} - sleepwalking through the development and deployment of technology without reflection. Yet in the general culture of ML research, this slumber has morphed into technological gaslighting ; the field is aware its creations rarely live up to or often fail their promises, yet it perpetuates unwarranted hype, voicing no concerns while reaping the rewards of inflated expectations. Bruno Latour reminds us in \textit{Love Your Monsters} (2012) that sin is not in creation but in abandonment. ML scientists produce “context-free” models - but fail to care for the social worlds their knowledge reorders. The goal of ML research for societal impact and meaningful innovation can benefit from accountability for social claims, not evading them. On that note, we offer \textit{articulation and justification of social claims} as one of many potential ways for ML research to evolve as situated actors who love and care for their knowledge products.

\section*{Acknowledgements}
In addition to his co-authors, Tianqi is especially grateful to his mentors: Daniel Susser, for providing feedback on an earlier draft of this work; David Gray Widder, for supporting his line of work and career development; and Leif Hancox-Li, for their generous commitment to improving Tianqi’s other writings and guiding his PhD studies.

\bibliography{aaai25}

\end{document}